\documentstyle[11pt]{elsart}    
\def\ltsim{\raise 2pt \hbox {$<$} \kern-1.1em \lower 4pt \hbox {$\sim$}}
\def\ltapprox{\raise 2pt \hbox {$<$} \kern-1.1em \lower 5pt \hbox {$\approx$}}
\def\gtsim{\raise 2pt \hbox {$>$} \kern-1.1em \lower 4pt \hbox {$\sim$}}
\def\gtapprox{\raise 2pt \hbox {$>$} \kern-1.1em \lower 5pt \hbox {$\approx$}}
\def\arcsec{$^{\prime\prime}$}
\def\arcmin{$^{\prime}$}
\def\degrees{$^{\circ}$}
\def\etal{{\it et al.~}}

\def\skuno{\vskip 20pt}

\def\p0{\phantom{0}}

\def\ph1{\phantom{1}}
\begin{document}
\begin{frontmatter}

\title
{Radio halo and relic candidates from the NRAO VLA Sky Survey}
\skuno
\skuno
\author[fis]{G. Giovannini\thanksref{g}},
\author[ira]{M. Tordi\thanksref{m}} and
\author[ira]{L. Feretti\thanksref{l}}

\address[fis]{
Istituto di Radioastronomia -- CNR, via Gobetti 101, I--40129
Bologna, Italy.
Dip. Fisica, Univ. Bologna, 
Via Berti-Pichat 6/2, I--40127 Bologna, Italy.}
\address[ira]{
Istituto di Radioastronomia -- CNR, via Gobetti 101, I--40129
Bologna, Italy.}

\thanks[g]{ggiovann@ira.bo.cnr.it}
\thanks[m]{mtordi@ira.bo.cnr.it}
\thanks[l]{lferetti@ira.bo.cnr.it}

\begin{abstract}

We present the first results of the search of
new halo and relic candidates in the NRAO VLA Sky Survey. We have
inspected a sample of 205 clusters from the X-ray-brightest Abell-type
clusters presented by Ebeling \etal (1996), and found 29 candidates.
Out of them, 11 clusters are already known from the literature to contain
a diffuse cluster-wide source, while in 18 clusters this is the first
indication of the existence of this type of sources. We classify these
sources as halos or relics according to their location in the cluster
center or periphery, respectively.
We find that the occurrence of cluster halos and relics is higher in 
clusters with high X-ray luminosity and high temperature. We also confirm
the correlation between the absence of a cooling flow and the presence of a 
radio halo at the cluster center.


\end{abstract}


\end{frontmatter}

\vfill\eject
\section{Introduction}

Large-scale  radio halos  in clusters of galaxies are  diffuse
radio sources with no apparent parent galaxy, typical sizes of 1 Mpc,
low surface brightness and steep radio spectrum.  They 
demonstrate the existence of relativistic electrons and large scale 
magnetic fields in the intracluster medium. 
The sources classified as {\it radio halos} are located at the cluster 
centers. Sources with similar properties have also been found
at the cluster peripheries: they are called {\it relic} sources. 
Moreover, in some clusters with a central dominant galaxy, the
relativistic particles can be traced out quite far, forming what is called a
{\it mini-halo} (see e.g. 3C~84 in the Perseus cluster, Burns \etal 1992). 

The radio halos and relics are a rare phenomenon. They are present in a
few  rich clusters, characterized by  high X-ray luminosity and high
temperature.  
The prototypical example of cluster-wide  radio halo is Coma-C, in Coma
(see Giovannini \etal 1993 and
references therein). The Coma cluster also contains the peripheral relic
1253+275, which
is connected to Coma-C through a very low-brightness bridge of radio
emission  (see e.g. Feretti \& Giovannini 1998). 
Other radio halos studied in detail so far are  those in A~2255 
(Burns \etal 1995, Feretti \etal 1997a), and in A~2319 (Feretti \etal 
1997b).  Well studied  relic sources are those 
 in A~2256 (R\"ottgering \etal 1994), 
A~3667 (R\"ottgering \etal 1997),  and A~85 (Bagchi \etal 1998).

The properties of halos and relics are not yet well understood,
because of the low number of known sources of this type. 
Also, it is not yet clear if radio halos
and relics have a common origin and evolution, or should be considered 
as different classes of sources. 
Information on a larger sample of halos and relics is crucial to investigate
their formation and evolution, and their relation to other cluster
properties.
To this aim we undertook the search for new candidates of radio halos
and relics in the NRAO\footnote {The National Radio Astronomy 
Observatory is operated by Associated Universities, Inc., under contract 
with the National Science Foundation.}
 VLA Sky Survey (NVSS, Condon \etal 1998). 
In this paper we report the detection of new halo and relic candidates,
which significantly increase the number of diffuse sources in clusters. 
The paper is organized as follows:  in Sect. 2 we describe the sample of 
clusters which were inspected in the NVSS, in Sect. 3 we present the results,
in Sect. 4 we give comments on the individual sources and their clusters, 
in Sect. 5 we discuss the results.

A Hubble constant H$_0$=50 km s$^{-1}$ Mpc$^{-1}$ and a deceleration 
parameter q$_0$=1 are assumed throughout.

\section {Sample {\bf and Source} selection}

We searched for cluster-wide radio sources in clusters using the public images
of the NVSS. This radio survey  was performed 
at 1.4 GHz with the Very Large Array (VLA) in the tightest configuration (D).
It has an angular resolution of 45\arcsec~ (HPBW),
a noise level of 0.45 mJy (1$\sigma$) and covers
all the sky north of Declination = --40\degrees. 

As a cluster sample we used the sample of X-ray-brightest 
Abell-type clusters (XBACs) presented by
Ebeling \etal (1996), consisting of 283 clusters/subclusters from 
the catalogue of Abell, Corwin \& Olowin (1989, ACO) detected in the 
ROSAT All Sky survey (RASS) with X-ray flux f$_X>$ 5 10$^{-12}$
erg cm$^{-2}$ s$^{-1}$ in the 0.1-2.4 keV energy range.
This is an all-sky, X-ray flux limited sample, 
complete in the galactic latitude range $\mid b \mid
\ge$20\degrees~ and in the redshift interval z$\le$0.2, but 
it contains also  12 clusters at lower galactic latitude and 24 clusters 
with redshift greater than 0.2 that meet the flux criterion 
(see  Tables 3 and 4 in Ebeling \etal 1996).

Because of the lack of short  baselines, 
the NVSS is insensitive to extended structure larger than 15\arcmin.
Since the known radio halos are about 1 Mpc in size, sources of this
type are missed in the NVSS if belonging to nearby clusters. 
We note indeed that the radio halo in Coma (z = 0.0232)
is  resolved out in the NVSS, because of the lack of short spacings. 
We have taken 
as a limiting redshift for the search of diffuse cluster-wide halos and
relics the value z=0.044, which corresponds to a largest detectable linear
size of about 1 Mpc. Considering this redshift limit, 
and taking also into account the declination limit of 
the NVSS, we ended with a sample of 207 clusters. For each of them, we 
searched for the NVSS image, and extracted a field of 
1\degrees$\times$1\degrees, centered on the cluster position given 
in the XBACs catalogue. We remark here that this is the X-ray position
and  may be different from that of the optical cluster center.
Moreover, in the case that the X-ray structure is double or more 
complex, the XBACs position refers to the centroid of the X-ray emission.
The two clusters
A~1773 and A~3888 fall in the few remaining gaps of the NVSS, and therefore
no image is currently available for them. For a few other clusters,
the field retrieved from the NVSS is slightly smaller than 1 square
degree, because of the existence of gaps at one of the field edges; however,
the available image covers a significant region of the clusters.
In conclusion, we have finally inspected  205 clusters, 
for a total of about 0.062 steradians. The  redshift distribution 
of the searched clusters with respect to the distribution of the 
whole XBACs catalogue is shown in Fig. 1.

\begin{figure}
\includegraphics{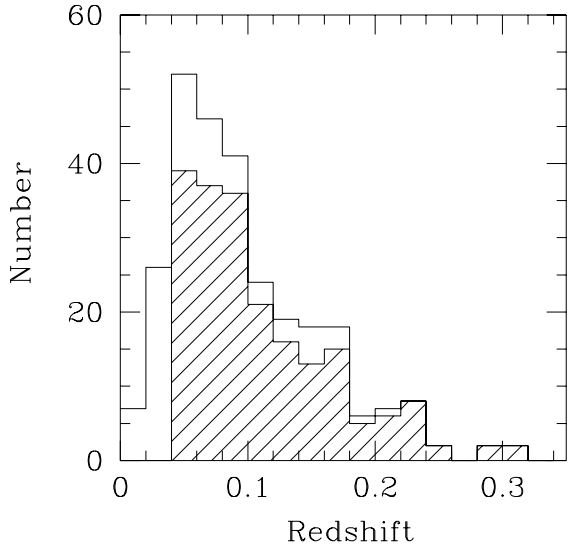}
\vspace{8 cm}
\caption{
Redshift distribution of the total sample of the XBACs, with the
shaded area representing the clusters inspected
in the NVSS.} 
\label{}
\end{figure}

To search for the cluster diffuse sources, we used the NVSS images 
overlaid upon the
optical images from the Digitized Palomar Sky Survey (PSS). 
We considered as diffuse cluster sources the radio features with surface
brightness greater than the 3 $\sigma$ level, which were found to be:
 i) resolved, ii) not
associated with bright galaxies, iii) not clearly related to extended
radio galaxies, iv) not simply attributable to the blend of
pointlike sources.
Thanks to the large images retrieved from the NVSS, we are confident 
that we could easily recognize and avoid the side lobes of near strong sources 
possibly present because of dynamical range problems.

We looked  into the literature and in the VLA Faint Images of the Radio Sky 
at Twenty-centimeters (FIRST) Survey (Becker \etal 1995)  
to discriminate the diffuse sources from discrete 
unrelated radio sources.
In some cases, where no high resolution data are available and the
radio structure is ambiguous, 
we considered the halo and relic candidates as uncertain.
We included in our sample also a few cases where the evidence 
of an extended radio emission is marginal 
but the presence of a diffuse emission was already known from the 
literature (see notes to the individual clusters). 

\section {Results}

With the above criteria, we have found 
diffuse extended emission in 29 clusters, listed in Table 1.
In 9 clusters, the presence of a diffuse extended radio source is  uncertain, 
because of the possible contamination of discrete radio galaxies or
because the diffuse radio emission is very faint. These clusters are
indicated by ``u'' in column 7 of Table 1.

\begin{figure}
\includegraphics{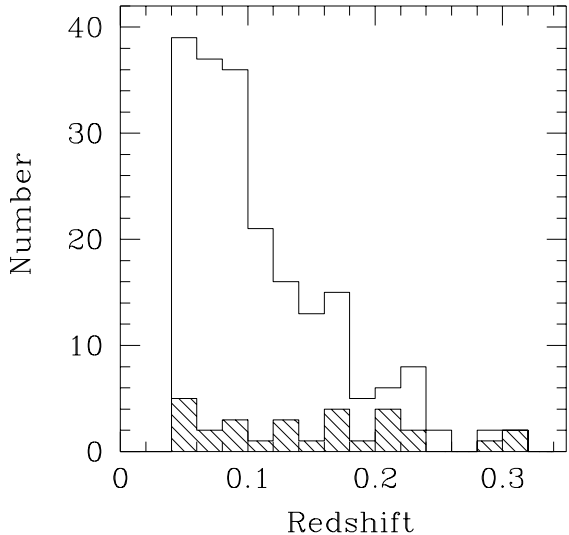}
\vspace{8 cm}
\caption{
Redshift distributions of the clusters with NVSS image and 
of the clusters with a diffuse source (shaded).}
\label{}
\end{figure}

The redshift distribution of clusters with diffuse sources with respect to 
that of the inspected clusters is presented in Fig. 2. 
We remind that we could have missed extended sources because
of their low surface brightness. We also
are aware that we can miss very extended sources ($>$ 1 Mpc) in the
clusters at lowest redshifts because of the poor sampling of short
baselines. This implies that the sample of diffuse sources found in this
search might be slightly biased toward sources at high redshift. 
On the other hand, we 
note that in distant clusters diffuse extended sources could be, 
in some cases, considered as discrete sources because of the large beam.

According to the literature, we classify the detected extended
source as halos, if they are centrally located in the cluster, or 
relics if they are peripheral. In the case that peripheral relics
are   projected onto the cluster center, they would be classified
as halo sources. More detailed observations will be necessary for
a correct classification.
The list presented in Table 1 contains most of the well known 
radio halos and relics, but also includes 18 clusters
where this is the first indication of the existence of a diffuse
source (see last column in Table 1). Among the latter, 11 are considered 
good candidates, while 7 are uncertain. 

The images of all the diffuse radio sources, overlaid onto the red 
images of the PSS, are given in Fig. 3. 

We present in Table 2 the parameters of the halo and relic candidates,
excluding the uncertain diffuse sources, for which 
the measurement of flux density and extension is very difficult 
because  strong contaminating sources are often present.
The radio flux density refers to the extended
emission after subtraction of obvious discrete sources. The size
is the largest dimension of the diffuse radio emission. 
We are aware that flux densities and sizes
 can be quite underestimated, because
of the low sensitivity to extended structures. Nevertheless, 
more than half of  the sources given in Table 2 
have a size larger than 1 Mpc. For the relics, we also give
the projected distance between the approximate centroid of the radio emitting
region and the X-ray cluster center.

\begin{table*}
\caption{List of cluster with halos or relics}
\begin{flushleft}
\begin{tabular}{llllllll}
\hline 
\noalign{\smallskip}
Name &  z & RA (J2000) & DEC  & T & L$_X$(0.1-2.4) & Radio Type  & Previous \\
     &    & h~~m~~s  & \degrees~~\arcmin~~\arcsec & keV 
& 10$^{44}$ erg s$^{-1}$   \\
\noalign{\smallskip}
\hline
\noalign{\smallskip}
 A~13  & 0.0943$^M$  &  00 13 32.2 & --19 30 03.6  &  4.3$^e$  &  2.24 & R 
& n \\ 
 A~2744  & 0.3080   &  00 14 16.1 & --30 22 58.8 & 11.04$^{AF}$ & 22.05 & H &
n \\
 A~22    & 0.1310   &  00 20 38.6 & --25 43 19.2  & 6.3$^e$   & 5.31 & u &
n  \\
 A~85    &  0.0555$^P$  &  00 41 48.7 & --09 19 04.8  &  6.2   &  8.38 & R &
y \\
 A~115   &  0.1971  &  00 55 59.8 & +26 22 40.8  &  9.8$^e$  & 14.57 & R & n
\\
 A~133  &  0.0603  &  01 02 45.1 & --21 52 48.0  &  3.8   &  3.57 & u & n \\
 A~209   &  0.2060  &  01 31 50.9 & --13 36 28.8  &  9.6$^e$  & 13.75 & u &
n \\
 A~401   &  0.0739  &  02 58 56.9 & +13 34 22.8  &  7.8   &  9.88 & H & y
\\
 A~520   &  0.2030  &  04 54 07.4 & +02 55 12.0  & 8.33$^{AF}$  & 14.20 & H &
 n \\
 A~545   &  0.1540  &  05 32 23.3 & --11 32 09.6  &  5.5   &  9.29 & H & n \\
 A~548b  &  0.0424$^{DHK}$  &  05 45 27.8 & --25 54 21.6 & 2.4 & 0.30 & R & n 
\\
 A~665   &  0.1818  &  08 30 57.4 & +65 51 14.4  & 9.03$^{AF}$  & 16.22 & H 
& y\\
 A~754   &  0.0542  &  09 09 01.4 & --09 39 18.0  &  8.7   &  8.01 & u & y
\\
 A~773   &  0.2170  &  09 17 54.0 & +51 42 57.6  & 9.29$^{AF}$  & 12.52 & H &
n \\
 A~1300  &  0.3071$^L$ &  11 31 54.9 & --19 54 50.4  & 5$^L$  & 23.40 & H & y
\\
 A~1664  &  0.1276  &  13 03 44.2 & --24 15 21.6  & 6.5$^{A}$ &  5.36 & R & n
\\
 A~1758a &  0.2800  &  13 32 45.3 & +50 32 52.8  &  8.7$^e$  & 11.22 & u & n
\\
 A~1914  &  0.1712  &  14 26 02.2 & +37 50 06.0  & 10.7$^e$  & 17.93 & H & n
\\
 A~2069  &  0.1145  &  15 24 09.8 & +29 55 15.6  &  7.8$^e$  &  8.74 & u & n
\\
 A~2142  &  0.0894  &  15 58 22.1 & +27 13 58.8  & 11.0   & 20.74 & u & y
\\
 A~2163  &  0.2080  &  16 15 49.4 & --06 09 00  & 13.83$^{AF}$ & 37.50 & H & y
\\
 A~2218  &  0.1710  &  16 35 52.8 & +66 12 50.4  & 7.05$^{AF}$ &  8.99 & H & y
\\
 A~2219  &  0.2281  &  16 40 22.5 & +46 42 21.6  & 12.42$^{AF}$ & 19.80 & u & n
\\
 A~2256  &  0.0581  &  17 04 02.4 & +78 37 55.2  &  7.5   &  7.05 & R & y \\
 A~2255  &  0.0809  &  17 12 45.1 & +64 03 43.2   & 7.3    & 4.79 & H & y \\
 A~2254  &  0.1780  &  17 17 46.8 & +19 40 48.0  &  7.2$^e$  &  7.19 & H & n \\
 A~2319  &  0.0555  &  19 21 05.8 & +43 57 50.4  & 9.3$^{AF}$   & 13.71 & H &
y \\
 A~2345  &  0.1760  &  21 26 58.6 & --12 08 27.6  &  8.2$^e$  &  9.93 & R+R & n
\\
 A~2390  &  0.2329  &  21 53 36.7 & +17 41 32.2 1 & 10.13$^{AF}$ & 21.25 & u
 & n \\
\noalign{\smallskip}
\hline
\label{olog}
\end{tabular}
\end{flushleft}
\end{table*}

\vfill\eject

Caption. Col. 1: cluster name; Col. 2: redshift;
Cols. 3 and 4: coordinates of the X-ray cluster center; 
Col. 5: temperature, where ``e'' indicates that the temperature has been
estimated from the L$_X$-kT relation;
Col. 6: X-ray luminosity in the ROSAT band (0.1-2.4 keV);
Col. 7: type of the diffuse radio source (H = halo, R = relic, u = uncertain);
Col. 8: previous knowledge in the literature of the existence of a 
diffuse source in this cluster (n = no; y = yes, reference given in Sect. 4).
\par
The data in Cols. 2, 3, 4, 5 and 6 are taken from Ebeling \etal 
(1996), except where a more recent reference is given. 
References are as follows: A = Allen \etal 1995;
AF: Allen \& Fabian 1998; DHK = Den Hartog \& Katgert 1996; 
L = L\'emonon \etal 1997; M = Mazure \etal 1996; 
P = Pislar \etal 1997.
\vfill\eject

\begin{figure}
\vspace{6 cm}
\caption{
Radio images of the diffuse sources (contours), overlaid onto the optical 
image from the PSS (grey-scale). Contour levels are 0.9, 1.35, 
2, 4, 8, 16, 32, 64, 128, 256 mJy/beam.}
\label{}
\end{figure}







\begin{table}
\caption{Parameters of the halo and relic candidates}
\begin{flushleft}
\begin{tabular}{llllll}
\hline 
\noalign{\smallskip}
Name &  Flux & $\theta$ & LLS & Dist. & Power \\
     &  mJy  & \arcmin & kpc & kpc & 10$^{24}$ W Hz$^{-1}$ \\
\noalign{\smallskip}
\hline
\noalign{\smallskip}
 A~13    & 34 & 6.4 &  880 & 150 & 1.30 \\
 A~2744  & 38 & 5.4 & 1700 & -- & 15.5 \\
 A~85    & 46 & 5.5 & 480  & 530 & 0.61 \\
 A~115   & 80 & 6.2 & 1500 & 1050 & 1.34 \\
 A~401   & 25 & 5.3 & 590  & -- & 0.59 \\
 A~520   & 38 & 4.4 & 1080 & -- & 6.74 \\
 A~545   & 41 & 7.4 & 1500 & -- & 4.18 \\ 
 A~548b  & 50 & 5.3 & 360  & 620 & 0.39 \\
 A~665   & 31 & 4.8 & 1100 & -- & 4.41 \\
 A~773   & 14 & 3.1 & 800  & -- & 2.84 \\
 A~1300  &  14 & 2.5 & 780 & -- & 5.7 \\
 A~1664  & 107 & 8.0 & 1400 & 1350 & 7.50 \\
 A~1914  & 50 & 4.4 & 960  & -- & 6.31 \\
 A~2163  & 55 & 6.0 & 1500 & -- & 10.2 \\
 A~2218  & 9  & 2.3 & 510  & -- & 1.13 \\
 A~2256  & 397 & 12.1 & 1100 & 590 & 5.77 \\
 A~2255  & 45  & 5.2  & 630  & --  &  1.27 \\
 A~2254  & 32 & 5.1 & 1140 & -- &  4.36 \\
 A~2319  & 23 & 4.8 & 420 & -- &  0.30 \\
 A~2345  & 92 & 5.4 & 1200 & 910 &  12.3 \\
         & 69 & 7.0 & 1560 & 2050 &  9.20 \\
\noalign{\smallskip}
\hline
\label{olog}
\end{tabular}
\end{flushleft}
\par\noindent
Caption. Col. 1: cluster name; Col. 2: Flux density at 1.4 GHz, after 
subtraction of obvious discrete sources; 
Col. 3: maximum angular size;  
Col. 4: largest linear size; 
Col. 5: approximate distance from the cluster center, in the case of relics;
Col. 6: monochromatic radio power at 1.4 GHz.
\end{table}

\section {Individual sources} 

In the following, we give comments on the individual sources and
discuss their reliability. 
\par\noindent
{\it A~13}. The diffuse radio source is not much displaced from the
cluster center, however it is elongated in shape and
does not include the central cluster galaxies. According to the data 
reported by Slee \etal (1996), the radio sources identified with
cluster galaxies account for a total flux of 3.9 mJy, confirming 
the presence of  extended radio emission.
\par\noindent
{\it A~2744}. Beside the centrally located radio emission, there is
also extended structure toward NE which could be either a relic or 
still related to the halo. 
\par\noindent
{\it A~22}. The diffuse source could be the due to a Wide Angle Tailed
radio galaxy plus a Narrow Angle Tailed radio galaxy. It is located
at about 14\arcmin~ from the cluster center, corresponding to 
$\sim$2.5 Mpc and would therefore be a relic.
\par\noindent
{\it A~85}. The relic in this cluster has been recently studied by Bagchi
 \etal (1998).
\par\noindent
{\it A~115}. This cluster shows a double morphology in X-ray. The relic
belongs to the northern clump.
\par\noindent
{\it A~133}. Komissarov and Gubanov (1994) report the existence of a very 
steep spectrum radio source ($\alpha >$ 2) coincident with the first
ranked galaxy of this cluster. The high resolution image published by 
Slee \etal (1994) shows a northern diffuse component of $\sim$100
kpc, probably not related to the radio galaxy.
The emission detected in the NVSS 
extends to the South where another discrete radio source is present
(see also the image by Owen \etal 1993). The existence of a  
cluster-wide diffuse emission is uncertain, as well as its connection
to the steep spectrum northern diffuse component.
\par\noindent
{\it A~209}. The presence of extended emission is uncertain due to the 
existence of strong discrete sources.
\par\noindent
{\it A~401}. The diffuse emission is very faint and located around the 
central cD galaxy, unlike the previous images by Harris \etal (1980a)
and Roland \etal (1981).
\par\noindent
{\it A~520}. The diffuse emission is centrally located, but of 
irregular shape with the present sensitivity.
\par\noindent
{\it A~545}. Despite of the presence of a strong discrete source at the
cluster center, the diffuse emission is easily visible. It is
rather symmetric and centrally located.
\par\noindent
{\it A~548b}. This cluster consists of many X-ray subclumps (Davis \etal 
1995). We detect a diffuse source classified as a relic in the NW cluster
region, but also extended emission is present around a radio galaxy
to the North. In the following analysis, we only consider the NW relic.
Further more sensitive observations should clarify whether 
the two extended features are bright regions of the same source.
\par\noindent
{\it A~665}. The presence of a halo was first reported by Moffet
\& Birkinshaw (1989) and confirmed by Jones \& Saunders (1996). 
\par\noindent
{\it A~754}. The existence of a halo in the center of this cluster was 
suggested by Harris \etal (1980b). 
At the location of the previously reported halo, 
there are many cluster galaxies, which could account for the emission.
Some diffuse emission of size $\sim$250 kpc is also detected in the NVSS
in the eastern peripheral region. The overall cluster-wide extended 
emission in this cluster is considered uncertain.  
\par\noindent
{\it A~773}. The diffuse emission is rather regular in shape
and centrally located. The FIRST image shows 3 discrete radio 
sources whose total flux density is lower that that detected in the NVSS,
confirming the presence of diffuse structure.
\par\noindent
{\it A~1300}. The presence of a central radio
halo is reported by Lmonon \etal (1998), who also classify the SW emission
as a relic. Since from the NVSS map the relic is poorly resolved,
we only consider in this paper the central radio halo.
\par\noindent
{\it A~1664}. 
The diffuse radio emission is located in the
SW peripheral region of the cluster and shows 
a regular structure, unlike the relics in Coma and A~3667, which
are generally elongated. The X-ray brightness distribution in this cluster 
is centrally peaked, with an asymmetric extension in the
direction of the diffuse radio source (Allen \etal 1995).
\par\noindent
{\it A~1758a}. The strongest source to the S is identified with a
cluster galaxy. It shows a Narrow Angle Tailed structure (O'Dea \& Owen 1985)
with the tail oriented to the SE. In the FIRST image, the easternmost 
structure is resolved in two compact sources, while the extended emission
in between is not detected. It could be a faint halo or the blend
of individual radio sources.
\par\noindent
{\it A~1914}. A very steep spectrum radio source ($\alpha >$2)
is reported by Komissarov \& Gubanov (1994). From higher resolution
images (Roland \etal 1985  and the image retrieved
from the FIRST survey) it is evident that 
the discrete sources cannot account for the extended emission.
\par\noindent
{\it A~2069}. The diffuse emission is located to the SE with respect  to
the cluster center at a distance of $\sim$6 Mpc, therefore its
classification as a cluster relic is uncertain. The image of
this region retrieved from the FIRST survey shows a faint point-like source
coincident with the southernmost peak, while the whole structure is
resolved out.
\par\noindent
{\it A~2142}. The presence of a halo in this cluster was suggested 
by Harris \etal (1977). The image here shows that the diffuse radio emission
is $\sim$350 kpc in size and is  located around a cluster galaxy. 
Therefore, we still consider uncertain the classification of this 
source as a cluster-wide radio halo.
\par\noindent
{\it A~2163}. The presence of a powerful and very extended
radio halo has been reported by Herbig \& Birkinshaw (1994).
\par\noindent
{\it A~2218}. The existence of a small radio halo was reported by
Moffet \& Birkinshaw (1989). Here, the diffuse source is barely 
visible to the SW of the strongest radio source.
\par\noindent
{\it A~2219}.  The strong radio source at the cluster center has a tailed
structure (as detected in the FIRST survey). It is 
difficult to safely establish the existence of a diffuse radio emission.
\par\noindent
{\it A~2256}. The extended emission detected here is the brightest region of 
the complex diffuse radio source in A~2256. This region studied  by
R\"ottgering \etal (1994) is at the cluster periphery,  while 
a very low brightness emission, not visible here,  permeates the cluster 
center as detected at 610 MHz by Bridle \& Fomalont (1976).
\par\noindent
{\it A~2255}. This cluster contains a central radio halo and a peripheral
relic (Burns \etal 1995, Feretti \etal 1997a). 
Both features are visible in the NVSS, although very faint and at the 
limit of significance. Based only on 
the NVSS image, the existence of  diffuse sources in this cluster
would be considered very uncertain. We note, however, that the radio halo
in A~2255 is best imaged at 90 cm, while at 20 cm  the halo structure 
is spotty and irregular also with higher sensitivity observations 
(Feretti \etal 1997). 
\par\noindent
{\it A~2254}. The extended emission is centrally located and shows a regular
structure. 
\par\noindent
{\it A~2319}. This radio halo has been studied in detail by Feretti \etal (1997b).
\par\noindent
{\it A~2345}. There are two peripheral extended sources in this cluster
located approximately on opposite sides with respect to the cluster 
center, at distance of $\sim$0.9 and $\sim$2 Mpc.
 If both sources will be confirmed as diffuse relics,
this cluster will be quite peculiar. The only known cluster with 
2 relics is A~3667  (R\"ottgering \etal 1997).
\par\noindent
{\it A~2390}. The extended emission in this cluster is uncertain, because
of the presence of a discrete strong source (Owen \etal 1993). 

\section {Discussion}

The number of new diffuse halos and relics detected in the NVSS is rather 
high, especially considering that these sources are characterized by steep 
spectrum and therefore they are better imaged at frequencies lower
than 1.4 GHz, and also taking into account the limited surface brightness
sensitivity. 

\begin{figure}
\includegraphics{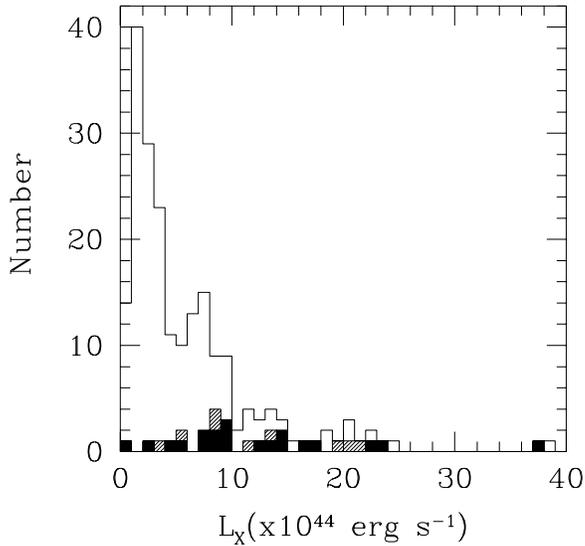}
\vspace{8 cm}
\caption{Distribution of the X-ray luminosity of the clusters 
searched for the presence
of a diffuse source. The black squares indicate the halo and relic 
candidates, while the
dashed squares indicate the uncertain sources (see Table 1). 
}
\label{}
\end{figure}

The percentage of clusters showing diffuse sources is higher in
clusters with high X-ray luminosity as can be deduced from Fig. 4 and
Table 3.
This trend can also be inferred from Fig. 2, due to the fact that the
XBACs sample is flux limited.
The percentage of clusters with diffuse sources is in the range 6\% to 9\%
in the clusters with L$_X \le$ 10$^{45}$ erg s$^{-1}$ and becomes
27\% to 44\% in clusters with L$_X >$ 10$^{45}$ erg s$^{-1}$. 
This effect could partly reflect the mentioned bias in the
sample against the detection of very extended diffuse  sources in the nearby 
clusters. However, given that a lower limit to the redshift 
is assumed, the lack of diffuse sources in low X-ray luminosity
clusters is real. This result is in agreement with
previous findings that cluster-wide
halos are present in massive clusters with high X-ray luminosity and 
high temperature (see e.g. Feretti \& Giovannini 1996).
Therefore, the diffuse cluster sources are not a rare phenomenon if X-ray 
luminous clusters are considered.

\begin{table}
\caption{Occurrence of diffuse cluster sources}
\begin{flushleft}
\begin{tabular}{llll}
\hline 
\noalign{\smallskip}
Lum. range &  N$_s$ & N$_f$ & N$_u$ \\
  10$^{44}$ erg s$^{-1}$    \\
\noalign{\smallskip}
\hline
\noalign{\smallskip}
 L$_X \le$ 10 & 173 & 11 & 4 \\
 10 $<$ L$_X \le$ 20 & 23 & 6 & 3 \\
 20 $<$ L$_X \le$ 30 & 7 & 2 & 2 \\
  L$_X >$30 & 2 & 1 & 0 \\
\noalign{\smallskip}
\hline
\label{olog}
\end{tabular}
\end{flushleft}
Caption. Col 1: X-ray luminosity range; Col 2: number of searched clusters;
Col 3: number of clusters with diffuse halos or relics; 
Col 4: number of clusters with uncertain diffuse sources.
\end{table}

\begin{figure}
\includegraphics{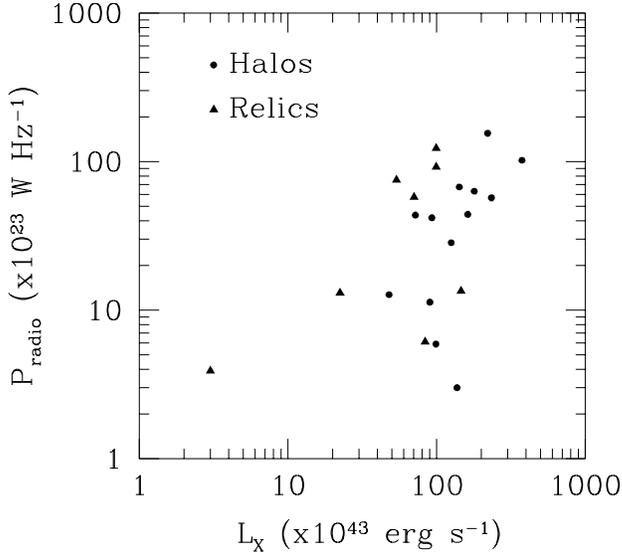}
\vspace{8 cm}
\caption{Plot of monochromatic radio power at 1.4 GHz versus the
cluster X-ray luminosity, for the halo and relic candidates.
}
\label{}
\end{figure}

We have looked for a possible correlation between the radio power of diffuse
sources and the cluster X-ray luminosity. The plot of these parameters
is given in Fig. 5. Although the data show a large scatter, the lack of high
power radio sources in clusters of low X-ray luminosity is clear, while
highly luminous X-ray clusters may have extended sources of either high 
or low radio power. Since the radio powers computed here may
be largely underestimated because of missing flux, due to the steep
spectrum of halos and relics and the poor sampling of short spacings,
further investigation of this correlation is needed.

A complete analysis of 
the connection between cluster properties and presence of diffuse
sources is beyond the scope of this paper. 
We only note that the clusters A~2744, A~520, A~545, A~773, and A~2254,
which are found here for the first time to host a central radio halo,
are all non-cooling flow clusters (White \etal 1997, Allen \& Fabian 1998).
Likewise, all the Abell clusters listed by Allen \& Fabian 
(1998) as non-cooling flow clusters are found to host a radio halo, with the
exception of A~2219, where the presence of a halo in considered uncertain but 
is possible. These results reinforce the strong correlation between 
the absence of a cooling flow and the presence of a radio halo at the 
cluster center.

\section {Conclusions}

We have presented in this paper new halo and relic candidates, found
in the NVSS after inspection of a sample of clusters from the XBACs
catalogue. We found 29 candidates, which can be divided as follows:
\par\noindent
$\bullet$ 11 clusters were already known from the literature. Out of them, 
7 contain radio halos (A~401, A~665, A~1300, A~2163, A~2218, A~2255, A~2319),
2 contain relics (A~85, A~2256), while the remaining 2 (A~754, A~2142) 
cannot be confirmed on the basis of the NVSS images, and are therefore
still considered uncertain.
We note that in the clusters A~1300 and A~2255, also the existence of 
a peripheral relic is reported in the literature.
\par\noindent
$\bullet$ in 18 clusters, this is the first indication of the existence of
a diffuse extended source. Among them, we found 6 clusters with
halos (A~2744, A~520, A~545,
A~773, A~1914, A~2254), and 5 clusters with
relics (A~13, A~115, A~548b, A1664, A~2345). The
cluster A~2345 contains actually 2 relics. 
The 7 clusters A~22, A~133, A~209, A~1758a, A~2069, A~2219 and A~2390
contain possible diffuse sources, which we indicate as uncertain. 

We found that the percentage of clusters showing diffuse sources is higher 
in clusters with high X-ray luminosity, being 6\%-9\% in clusters
with L$_X\le$ 10$^{45}$ erg s$^{-1}$ and 
27\%-44\% in clusters with L$_X >$ 10$^{45}$ erg s$^{-1}$. Therefore, 
the diffuse cluster sources are not a rare phenomenon if X-ray luminous
clusters are considered.

We have found no correlation between the radio power of diffuse sources and the
cluster X-ray luminosity. However, we note that the radio powers
may be largely underestimated because of missing flux in the NVSS images.

Finally, we note that the large majority of clusters hosting a radio halo
do not contain a cooling flow. This reinforces the strong correlation 
between the absence of a cooling flow and the presence of a 
radio halo in clusters.

\begin{ack}

We thank Dr. W.D. Cotton for discussions and advice on the NVSS.
LF and GG acknowledge partial financial support from the Italian
Space Agency (ASI) and from the Italian MURST.

This research has made use of the NASA/IPAC Extragalactic Database (NED)
which is operated by the Jet Propulsion Laboratory, Caltech, under contract
with the National Aeronautics and Space Administration.

\end{ack}

\vskip 1 truecm

{\bf References}

\par\medskip\noindent
  Abell, G.O., Corwin, H.G., Olowin, R.P., 1989, ApJS 70, 1
\par\medskip\noindent
  Allen, S.W., Fabian, A.C., Edge, A.C., B\"ohringer,
H., White, D.A., 1995, MNRAS 275, 741
\par\medskip\noindent
  Allen, S.W., Fabian, A.C., 1998, MNRAS 297, L57
\par\medskip\noindent
  Bagchi, J., Pislar, V., Lima Neto, G.B, 1998, MNRAS 296, L23
\par\medskip\noindent
  Becker, R.H., White, R.L., Helfand, D.J.,  1995, ApJ 450, 559
\par\medskip\noindent
  Bridle, A.H., Fomalont, E.B., 1976, A\&A 52, 107
\par\medskip\noindent
  Burns, J.O., Roettiger, K., Pinkney, \etal, 1995, ApJ 446, 583
\par\medskip\noindent
  Burns, J.O., Sulkanen, M.E., Gisler, G.R., Perley, R.A., 
1992, ApJ 388, L49
\par\medskip\noindent
  Condon, J.J., Cotton, W.D., Greisen, E.W. Yin, Q.F., Perley, R.A., 
Taylor, G.B., Broderick, J.J., 1998, AJ 115, 1693 
\par\medskip\noindent
  Davis, D.S. Bird, C.M., Mushotzky, R.F. Odewahn, S.C., 1995,
ApJ 440, 48
\par\medskip\noindent
  Den Hartog, R., Katgert, P.,  1996, MNRAS 279, 349
\par\medskip\noindent
  Ebeling, H., Voges, W., B\"ohringer, H., Edge, A.C., Huchra, J.P., 
Briel, U.G., 1996, MNRAS 281, 799
\par\medskip\noindent
  Feretti, L., Giovannini, G., 1996, In: Extragalactic Radio
Sources, IAU Symp. 175,  Eds. R. Ekers, C. Fanti \& L. Padrielli, 
Kluwer Academic Publisher,  p. 333
\par\medskip\noindent
  Feretti,  L.,  B\"ohringer, H., Giovannini, G., Neumann D.,
 1997a, AA 317, 432
\par\medskip\noindent
  Feretti, L., Giovannini, G. B\"ohringer, H., 
1997b, New Astronomy, 2, 501
\par\medskip\noindent
  Feretti, L.,  Giovannini, G., 1998, in {\it Untangling Coma 
Berenices: A new view of an old Cluster}, Word Scientific Publishing Co 
Pte Ltd, p 123
\par\medskip\noindent
  Giovannini,  G., Feretti, L., Venturi, T., Kim, K.-T., 
Kronberg, P.P., 1993,  ApJ, 406, 399
\par\medskip\noindent
  Harris, D.E., Bahcall, N.A., Strom, R.G., 1977, A\&A 60, 27
\par\medskip\noindent
  Harris, D.E., Miley, G.K.,  1978, A\&AS 34, 117
\par\medskip\noindent
  Harris, D.E., Kapahi, V.K., Ekers, R.D., 1980a, A\&AS 39, 215
\par\medskip\noindent
  Harris, D. E., Pineda, F. J., Delvaille, J. P.,
Schnopper, H. W., Costain, C. H., Strom, R. G., 1980b, A\&A 90, 283 
\par\medskip\noindent
  Herbig, T., Birkinshaw, M., 1994, BAAS, Vol. 26, No 4, 1403
\par\medskip\noindent
  Jones, M., Saunders, R., 1996, in {\it R\"ontgenstrahlung from
the Universe}, H.U. Zimmermann, J.E. Tr\"umper \& H. Yorke Eds., MPE
Report 263, p. 553
\par\medskip\noindent
  Komissarov, S.S., Gubanov, A.G., 1994, A\&A 285, 27
\par\medskip\noindent
  L\'emonon, L., Pierre, M., Hunstead, R., \etal, 1997, A\&A 326, 34
\par\medskip\noindent
  Lmonon, \etal, 1998, astro-ph 9811467
\par\medskip\noindent
  Mazure, A., Katgert, P, Den Hartog, R., \etal, 1996, A\&A 310, 31
\par\medskip\noindent
  Moffet, A. T., Birkinshaw, M., 1989, AJ 98, 1148
\par\medskip\noindent
  O'Dea, C.P., Owen, F.N., 1985, AJ 90, 927
\par\medskip\noindent
  Owen, F.N., White, R.A., Ge, J-P., 1993, ApJS 87, 135
\par\medskip\noindent
  Pislar, V., Durret, F., Gerbal, D., Lima Neto, G.B., Slezak, E., 
1997, A\&A 322, 53
\par\medskip\noindent
  Roland, J. Sol, H. Pauliny-Toth, I. Witzel, A. 1981,
A\&A 100, 7
\par\medskip\noindent
  Roland, J., Hanish, R.J., V\'eron, P. Fomalont, E., 1985, 
A\&A 148, 323
\par\medskip\noindent
  R\"ottgering, H., Snellen, I., Miley, G., \etal, 1994,
 ApJ  436, 654
\par\medskip\noindent
  R\"ottgering, H.J.A., Wieringa, M.H., Hunstead, R.W, Ekers, R.D.,
 1997,  MNRAS 290, 577 
\par\medskip\noindent
  Slee, O.B., Roy, A.L., Savage, A,  1994, Aust. J. Phys. 47, 145
\par\medskip\noindent
  Slee, O.B., Roy, A.L., Andernach, H., 1996, Aust. J. Phys. 49, 977
\par\medskip\noindent
  White, D.A., Jones, C., Forman, W., 1997, MNRAS 292, 419

\end{document}